\documentclass[twocolumn,pra,aps,showpacs]{revtex4}
\usepackage{xspace,amsmath,amsfonts,amsthm,amssymb,amsbsy,graphicx,color}

\begin{document}

\newtheorem{theo}{Theorem}
\newtheorem{lemma}{Lemma} 

\title{Conditions for the Quantum to Classical Transition: Trajectories vs. Phase Space Distributions}

\author{Benjamin D. Greenbaum} 

\affiliation{Department of Physics, University of Massachusetts at Boston, 100 Morrissey Blvd, Boston, MA 02125, USA} 

\author{Kurt Jacobs}

\affiliation{Department of Physics, University of Massachusetts at Boston, 100 Morrissey Blvd, Boston, MA 02125, USA}

\author{Bala Sundaram}

\affiliation{Department of Physics, University of Massachusetts at Boston, 100 Morrissey Blvd, Boston, MA 02125, USA}

\begin{abstract}
We contrast two sets of conditions that govern the transition in which classical dynamics emerges from the evolution of a quantum system.  The first was derived by considering the trajectories seen by an observer (dubbed the ``strong'' transition) [Bhattacharya {\em et al.}, Phys. Rev. Lett. {\bf 85} 4852 (2000)], and the second by considering phase-space densities (the ``weak'' transition) [Greenbaum {\em et al.}, Chaos {\bf 15}, 033302 (2005)]. On the face of it these conditions appear rather different. We show, however, that in the semiclassical regime, in which the action of the system is large compared to $\hbar$, and the measurement noise is small, they both offer an essentially equivalent local picture.  Within this regime, the weak conditions dominate while in the opposite regime where the action is not much larger than $\hbar$, the strong conditions dominate. 
\end{abstract} 

\maketitle

\section{Introduction}

It has been established by a number of recent works that the act of continuously observing a quantum system is sufficient to induce a transition from quantum to classical dynamics, so long as the action of the system is sufficiently large and the measurement sufficiently strong~\cite{{Spiller94,Schack95,Brun96,Percival98,Percival98b,Bhattacharya00,Habib02,Bhattacharya03,Ghose03,Ghose04,Everitt05,Ghose05}}.  Under these conditions the quantum system remains well-localized in phase space, any noise introduced by the measurement is negligible, and the mean position and momentum of the quantum particle follow the smooth trajectories of classical mechanics. In particular, this approach provides a detailed understanding of how classical chaos emerges from quantum dynamics in the classical limit.  Measurement (or equivalently the extraction of information by an environment, whether explicitly observed or not) is essential for this process: closed quantum systems cannot exhibit chaos, as demonstrated by results such as the Koslov-Rice theorem~\cite{Kosloff81,Manz89} (reviews of this topic are given in~\cite{Khanna06,GreenbaumPhD}).  

Prior to this type of analysis, research on the quantum-to-classical
transition focused on phase-space distribution functions, rather than
observed trajectories. If the initial conditions for a classical
system are not known precisely, and it is not measured during its
evolution, then the state of the system is described by an ever
broadening probability density in phase space. The dynamics of this
density are given by the classical Liouville equation~\cite{Gold}. A
quantum analog of this phase-space distribution is the Wigner
function~\cite{Wigner32}. For a classically chaotic, one-dimensional,
time-dependent Hamiltonian, it was found that the interaction with a
large (Markovian) environment would transform the dynamics of the
Wigner function into that of the classical phase-space density, at
least under some circumstances~\cite{Habib98}. The study of the
quantum-to-classical transition for phase-space densities under
generic environmental interactions is often referred to as
``decoherence''~\cite{Zurek02,Pattanayak03}. Heuristic arguments were
devised to explain this phenomenon for classically chaotic
systems~\cite{Zurek02} although, due to the complexity of the quantum
and classical evolution equations for these systems, such arguments
are not easy to make precise. Nevertheless, the mechanisms,
valid in the semiclassical limit, by which the Wigner function
closely approximates the classical density for one-dimensional,
time-dependent, classically chaotic systems have recently been
reported~\cite{GreenbaumNew,Greenbaum05} which provides one focus for the present work. 

The two approaches to the quantum-to-classical transition for open systems, the trajectory-level method employing continuous measurement theory, and the distribution approach involving the Wigner function, in fact, may treat precisely the same physical situation. When a quantum system interacts with a Markovian environment, this environment continually carries away information about the system. If an observer chooses to measure this information, the resulting dynamics is described by the stochastic master equation of continuous quantum measurement theory~\cite{JacobsSteck06,Brun02}. If the observer does not make use of this information, then the equation reduces to an evolution equation for the Wigner function under a Markovian environment, as employed in the studies of decoherence.  Note that the act of observing the environment has no additional effect on the {\it system} than that already imposed by the environment. This is why the standard distribution-level description of an environmental interaction is given by averaging over all possible realizations of the underlying trajectories~\cite{foot1}. 

As a result, continuous measurement theory can be used mathematically as a way to analyze the behavior of the Wigner function in the presence of an environment. This is because the measurement equations correctly describe the Wigner function dynamics regardless of whether an observer happens to be ``actually'' monitoring the system or not.  Thus a continuously measured system behaving classically at the trajectory level should exhibit a corresponding Wigner function which reproduces the classical phase space density, though the converse need not be true. Namely, a density undergoing a noise induced transition may not have a smooth classical trajectory picture. 

While continuous measurement will explain the emergence of classical
motion at the level of phase-space densities, there are other relevant
questions regarding the relationship between the emergence of
classicality at the two levels, densities and trajectories. In this
paper, we address two of these. The first is to define more precisely the circumstances under which the emergence of classicality at one level effects emergence at the other. Specifically, since phase-space densities can converge {\em without} the underlying, observed trajectories having become classical, we ask under what conditions the emergence of a classical phase-space density {\em does imply} that an observer would see the classical trajectories. The second, related, question regards two sets of conditions that govern the emergence of classicality. The first, derived by Bhattacharya {\em et al.}~\cite{Bhattacharya00,Bhattacharya03}, provide conditions under which the observed trajectories of a quantum system will obey classical dynamics. The second, derived by Greenbaum {\em et al.}~\cite{Greenbaum05,GreenbaumNew} show how the Wigner function matches its classical counterpart. These two sets of conditions were derived in quite different ways, involving different concepts, and we wish to understand the relationship between them.  

In what follows we will refer to the emergence of classicality at the level of the phase-space densities as the {\em weak} quantum-to-classical transition (weak QCT), and the emergence at the level of observed trajectories as the {\em strong} QCT~\cite{foot2}. In the next section we summarize the arguments used to derive the conditions for the emergence of classicality in both the strong~\cite{Bhattacharya03} and weak~\cite{Greenbaum05,GreenbaumNew} cases and present a useful reformulation of the latter. In Section~\ref{sec::III} we analyze the relationship between the weak and strong transitions. In particular we explore the nature of the regime where the weak transition implies the strong as opposed to the one in which the weak QCT is satisfied but the strong is not. In Section~\ref{sec::IV} we present an alternative approach to deriving the conditions in which the weak transition implies the strong transition. This 
is subsumed by the condition derived in Section~\ref{sec::III}. In Section~\ref{sec::V} we conclude with a brief summary of the main results.  
 
\section{Inequalities Governing the Quantum-to-Classical Transition}  

\subsection{The Strong QCT}

In references~\cite{Bhattacharya00,Bhattacharya03} Bhattacharya {\em
  et al.} derived a set of approximate inequalities governing the
emergence of classical motion in an observed quantum system consisting
of a single particle.  These inequalities define the strong QCT as they delineate the conditions under which an observed single particle will follow a localized classical trajectory.  For purposes of succinctness, we will, therefore, refer to the inequalities derived by Bhattacharya {\em et al.}  as the {\em strong inequalities}, since they relate to the QCT in the strong sense.  Through the paper, we will also denote the expectation values for the momentum and position of the single particle system as $x$ and $p$. 

The classical Hamiltonian for the system at $(x,p)$ is
generally time-dependent and of the form
\begin{equation}
  H(x,p,t) = \frac{p^2}{2m} + V(x,t)  
\end{equation}
where $F(x,t) = -\partial V(x,t)/ \partial x$ is classical force, and,
as usual $m$, is the particle mass. For the remainder of the work, we will not explicitly denote the
time-dependence of functions of phase-space variables.  The first of the strong inequalities determine when the centroid of the wave-function will remain sufficiently localized that the centroid will obey classical mechanics, and is divided into two regimes. When the strength of the non-linearity, as measured by the magnitude of $\partial^2_x F(x)$, is small enough to satisfy 
\begin{equation}
  |\partial_x^2 F | \ll \frac{4 |F| \sqrt{ m |\partial_x F|}}{\hbar}
            \, , \label{locineq1}
\end{equation}
then the condition is 
\begin{equation}
   k \gg \left| \frac{\partial_x^2 F }{8F} \right| \sqrt{ \frac{|\partial_x F|}{2m}} \,.  \label{locineq2} 
\end{equation}
When the strength of the non-linearity violates Eq.(\ref{locineq1}), the condition becomes 
\begin{equation}
    k \gg \left( \frac{\partial_x^2 F}{ 8 F} \right)^2 \frac{2\hbar}{m}  \, .  \label{locineq3} 
\end{equation}
Here $k$ is the ``measurement strength'', which is the parameter that determines the rate at which the environment extracts information about the system~\cite{DJJ}. An example is given by the weak-coupling, high temperature limit of the Caldeira-Leggett master equation describing a single particle interacting with a thermal environment~\cite{GZbook}.  In this case $k = D/\hbar^2$, where $D$ is the rate of momentum diffusion due to the environment. Note that while $k$ is a constant, $F$ depends on $x$, and thus varies over the phase space of the system. Thus, the right hand side of the inequalities above are understood as being averaged over the phase space, weighted by the relative time the particle spends at each point. 

The inequalities as given in \cite{Bhattacharya03} also include a
dimensionless quantity $\eta$, referred to as the {\em measurement
  efficiency}, which is the fraction of the extracted information that
is actually obtained by the observer. When considering the measurement
analysis merely as a tool to derive results regarding the transition
in terms of the Wigner function, $\eta$ is irrelevant. Thus, in
comparing the strong inequalities with the weak
transition derived by Greenbaum {\em et al.}, we will always set
$\eta=1$, corresponding to the assumption that any observer has all
the available information. Choosing a smaller value of $\eta$ is
useful only when considering the behavior of observed trajectories in
particular physical situations where the information 
available to observers is limited by practical considerations. 

The second part of the strong inequalities gives the condition under which the noise in the observed trajectories is negligible, so that they follow the smooth classical evolution given by the Hamiltonian. This consists of two inequalities that must both be satisfied: 
\begin{equation}
	\frac{ 2|\partial_x F| }{\bar s } \ll \hbar k \ll 
   	    \frac{ |\partial_x F|\bar{s} } {4}
   \,,
  \label{lnineq}
\end{equation} 
Here $\bar s$ is a measure of the action of the system in units of $\hbar$. Specifically, $\bar{s}\equiv\mbox{min}(S/\hbar,S'/\hbar)$, where 
\begin{eqnarray}
   S & = & \frac{|p|^3}{8 m |F|}  \label{eq:s1} \\ 
   S' & = & \frac{m |F|^3}{|p| (\partial_x F)^2} 
\end{eqnarray}
Both $S$ and $S'$ are expressions involving the system parameters that have units of action. 

The strong inequalities are thus given by Eq.(\ref{locineq2}) or (\ref{locineq3}), and Eq.(\ref{lnineq}). The first two state that the measurement must be strong enough to successfully limit the spreading of the wave-packet induced by the non-linearity. The second set, given in Eq.(\ref{lnineq}), state, essentially, that the action of the system in units of $\hbar$ should be sufficiently large so that there is a value of $k$ that satisfies both inequalities. As the action of the system becomes very large compared to $\hbar$, then effectively {\em any} measurement strength will satisfy these inequalities, and this defines the classical limit. 

\subsection{The Weak QCT}

The conditions derived by Greenbaum {\em et al.}~\cite{GreenbaumNew, Greenbaum05}
give a time-scale for when a Wigner function for a quantum system
driven by environmental noise will agree with a noise-driven classical
phase-space density. Moreover, the weak QCT has two distinct regimes depending on the noise level: a small noise regime in which the transition occurs after  the classical structure evolution is in a global steady state and another in which the transition occurs locally while large structures are still forming.  We now reformulate the conditions in~\cite{GreenbaumNew} to obtain an expression for the measurement strength which separates these regimes, allowing comparison with the strong inequalities. 
We also extend the results by providing a weak inequality relevant to the strong QCT low-noise condition. 

The arguments devised in~\cite{GreenbaumNew} proceed in two parts. First, a purely classical relation is derived which gives the
phase-space length scale, $l(t^*)$, below which noise will prevent the
creation of fine structure in the classical phase-space density beyond a time $t^*$.  This is derived
by calculating two phase-space lengths, both of which are functions of
time, and equating them. These lengths are scaled so as to have units
of the square root of phase-space area. The first is the length over
which the noise destroys fine structure as a function of time, which
is given by $l_{cl}(t) = \sqrt{D t/(m\bar{\lambda})}$ where
$\bar{\lambda}$ is the usual classical Lyapunov exponent defined over the bounded phase space region. $l_{cl}$ clearly increases with time. The second length is the scale of the phase-space structures developed by the dynamics, $\delta = \sqrt{\xi A} e^{-\bar{\lambda} t}$, which decreases with time. The steady-state length scale $l$ is the point at which these two match. Equating $l_{cl}$ and $\delta$, we obtain an expression for the diffusion constant in terms of the length scale $l$. This is 
\begin{equation}
  D(l) \approx  \frac{2 m \bar{\lambda}^2 l^2}{\ln(\xi A/l^2)}   \label{weak1}
\end{equation}
where $A$ is the phase-space area accessible to the system, and $l$ is a length with units of $\sqrt{A}$. There is however, an ambiguity in the value of $\xi$. This comes from the expression for the length scale of the fine structure in the classical density. Its role is to set the scale of the structure in the density of the initial state. As a result, $\xi$ can be anywhere in the range $[1,A/\hbar]$: the lower bound corresponds to an initial state that is uniform over essentially all phase space, and the upper bound to an initial state that is confined to a single cell of area $\hbar$.  This upper bound comes from the fact that any quantum phase-space density is limited to fine structure on the order of $\hbar$, and there is therefore no point in considering initial classical densities with finer structure. In fact, due to the logarithm in the expression for $D(l)$, the ambiguity in $\xi$ can be dealt with quite easily. To do so we merely choose the upper or lower bound, whichever provides the most stringent condition. That is, we choose the value of $\xi$ so as to err on the safe side.   

The second step is deriving a condition under which noise is sufficient to wash out interference fringes on length scales below $l_{cl}$.  This condition defines the weak QCT. In~\cite{Greenbaum05,GreenbaumNew} semiclassical arguments are used to show that inference fringes are washed out on the length scale of $l_{qu}(t) = \hbar \sqrt{\bar{\lambda} m/(Dt)} = \hbar/l_{cl}(t)$. If we set $l_{qu}=l_{cl}$, then we obtain a simple condition purely in terms of $l_{cl}$:
\begin{equation}
  l_{qu}^2 \approx l_{cl}^2\gtrsim \hbar  \label{weak2} 
\end{equation}
The weak QCT will occur for distributions at a time, $t_{qc}\approx m\hbar\bar{\lambda}/D \label{t_qc}$.  By equating $l$ and $l_{qu}$ or, equivalently, $t^*$ and $t_{qc}$, we find the threshold between two distinct weak QCT regimes, which define whether the weak QCT occurs before classical structure growth terminates.  Interpreting this noise as coming from measurement we set $D=\hbar^2 k$ yielding   
\begin{equation}
  k_{crit} \approx  \frac{2 m \bar{\lambda}^2 }{\hbar \ln(\xi A/\hbar)}   \label{weak3}.
\end{equation}
Further, we can identify $A$ with the action of the system, so that $\tilde{s} \equiv A/\hbar$ is an action for the system in units of $\hbar$. This gives 
\begin{equation}
  k_{crit} \approx  \frac{2 m \bar{\lambda}^2}{\hbar \ln(\xi \tilde{s})}   \label{weak4}
\end{equation}
Now, since we know that $\xi \in [1,\tilde{s}]$, we see that the difference between taking the maximum and minimum values of $\xi$ only results in a factor of two difference in the right hand side. To obtain our final expression, we take $\xi$ to have its minimum value as this results in the most stringent condition. The result is 
\begin{equation}
  k_{crit} \approx  \frac{2 m \bar{\lambda}^2}{\hbar \ln(\tilde{s})}   \label{weak5a}
\end{equation}
When $k$ is greater than this value the weak QCT will occur while classical structures continue to evolve, while for smaller values classical structures will stop forming before the weak QCT.  We also want a condition under which  noise is negligible so as to obtain the classical limit in the narrow sense. This will be true if the ``smearing area'' $l^2$ is small compared to the accessible phase space $A$. Imposing this condition on the relation in Eq.(\ref{weak1}), we have 
\begin{equation}
    k  \ll \left( \frac{m \bar{\lambda}^2 }{\hbar} \right) \frac{2\tilde{s}}{\ln(\tilde{s})}    \label{weak5b}
\end{equation} 
where this time we have set $\xi$ at its maximum value to obtain the most stringent condition. Putting the two inequalities together, we define the regime in which the weak QCT occurs while large classical structures continue to form
\begin{equation}
 \left( \frac{2 m \bar{\lambda}^2}{\hbar}\right) \left[  \frac{1}{\ln(\tilde{s})}\right] \; \lesssim \;  k \; \ll \; \left( \frac{2 m \bar{\lambda}^2 }{\hbar}\right) \left[ \frac{\tilde{s}}{\ln(\tilde{s})} \right]   \label{weak6}
\end{equation}
It is important to note that since the weak QCT has been understood using semiclassical arguments, we can only expect these arguments to be strictly valid in the semiclassical regime --- that is, when the dimensionless action of the system $s \gg 1$, and when the noise is relatively small in comparison to the classical dynamics (that is, when $l^2$ is small compared to the accessible phase-space area $A$).  

\section{The emergence of classicality: Weak vs. Strong}
\label{sec::III} 

We wish to examine the relationship between the weak and strong quantum-to-classical transitions. Unlike the weak QCT, the strong QCT only occurs {\it after} a minimum noise threshold is met.  The observed wave-function is highly localized in phase space and the noise on observed trajectories is negligible. In this case the weak QCT should also have taken place. That is, the quantum Wigner function will agree with the classical density, and this density will exhibit fine structure down to a length scale much smaller than the available phase space $A$. This result should follow immediately from the fact that 1) the Wigner function is merely the sum of the Wigner functions for all the possible localized observed wave-packets, 2) the centroid of each wave-packet obeys the classical equations of motion, and 3) each wave-packet has area $l^2$ and therefore has a width of order $l$ in each (dimensionless) phase-space direction.  

Secondly, if we are in the above highly localized regime, the weak QCT should imply the strong QCT. That is because the Wigner function would not exhibit the same fine structure (that is, the same structure of foliating unstable manifolds) as the classical density if the equivalent observed trajectories were not following the classical dynamics. (In fact, by considering the constraints on the trajectory Wigner functions implied by the scale of the fine structure, one can derive a quantitative condition for when the weak transition implies the strong, and we will do this in Section~\ref{sec::IV}.)

With the above discussion in mind, we now compare directly the strong and weak QCT.  This is easy to do if we approximate the local Lyapunov exponent by its global value.  This approach is consistent with the inequalities of Bhattacharya {\it et al.} in which one equates local forces with their phase-space averages. The local Lyapunov exponent measures the local stretching rate of a point in phase space, $(x_0, p_0)$.  The linearized Newton's equation for the perturbation, $\delta x$, then yields  
\begin{equation}
m\frac{d^2 \delta x}{dt^2}\approx \left. \partial_x F \right|_{x_0}\delta x.
\end{equation}  
The local Lyapunov exponent is defined by the solution to this equation:
\begin{equation}
\delta x(t)\approx \delta x_0 e^{\lambda t},
\end{equation}
where 
\begin{equation}
\lambda^2=\frac{|\partial_x F|}{m}.
\end{equation}
We now simply replace $\lambda$ with its average value over phase space, $\bar{\lambda}$ to complete the approximation. 

Using this relationship, the strong inequalities that give the conditions for low noise (Eq.(\ref{lnineq})) become 
\begin{equation}
  \left( \frac{ 2 m \bar\lambda^2 }{\hbar} \right) \left[ \frac{1}{\bar{s}} \right] \; \ll \;  k \; \ll \;    \left( \frac{ 2 m \bar\lambda^2 } { \hbar} \right) \left[ \frac{\bar{s}}{8} \right]. 
  \label{eq::strongln2}
\end{equation}
We see that these are very similar to the weak QCT regime defined by Eq.(\ref{weak6}). 

We assume that the ``actions'', $\bar s$ and $\tilde s$, that we associate with the system, are both approximately equal to the system action, and may therefore be equated. In the semiclassical regime, in which $\tilde{s} \gg 1$, we have both $\tilde{s} \gg \ln(\tilde s)$ and $\ln{\tilde s} \sim O(8)$, so that the weak regime above $k_{crit}$ and the strong low-noise criteria are essentially equivalent. This is logical, as the strong QCT assumes that the trajectories explore classical structures.  The caveats to this are that when $s$ is extremely large, being in this weak regime implies the strong low-noise inequality (both the left-hand inequality and the right-hand inequality). That is, the conditions for this regime are {\em stronger} than the strong low-noise inequality. By comparing Eqs. (\ref{eq::strongln2}) and (\ref{weak6}) we can write down a specific condition under which a system being in the $k>k_{crit}$ weak regime implies the strong low-noise inequality. This is 
\begin{equation}
   s \gg  e^8 \approx  3\times 10^{3} .   \label{eq::wtos0}
\end{equation}
In the opposite case, when $s$ is not much larger than unity, the strong low-noise inequality is satisfied over a range of $k$ values before $k_{cr}$ is reached signaling the start of the weak regime, though this requires relaxing the semiclassical condition, which may effect the validity of the weak approximation.

The above result raises a curious question. The derivation of the weak QCT above $k_{crit}$ would lead us to believe that this is a sufficient condition for the emergence of classical motion in the semiclassical regime defined by Eq.(\ref{eq::wtos0}), both at the trajectory and density levels.  However, the weak QCT is most easily compared to the strong inequalities that guarantee low noise (Eq.(\ref{lnineq})). The derivation of the strong inequalities implies that a second condition is required to guarantee classical behavior, this being the bound relating the noise to the size of the nonlinearity given either by Eq.(\ref{locineq2}) or Eq.(\ref{locineq3}). Either the weak QCT regime as derived is not as complete as previously assumed, or the part of the strong inequalities that bound the non-linearity is redundant in this semiclassical regime. 

It turns out that the answer is the latter.  That is, in the semiclassical regime defined by Eq.(\ref{eq::wtos0}), the weak QCT regime defined by Eq.(\ref{weak6}) also implies that both localization conditions (Eq.(\ref{locineq2}) and Eq.(\ref{locineq3})) are satisfied. To see this we note that it will be true if 
\begin{eqnarray}
    \frac{2m\bar{\lambda}^2}{\hbar \ln(s)} & \gg & \left| \frac{\partial_x^2 F }{8F} \right| \frac{\bar{\lambda}}{\sqrt{2}},  \label{raw1} 
\end{eqnarray}
and 
\begin{eqnarray}
    \frac{2m\bar{\lambda}^2}{\hbar \ln(s)} & \gg & \left( \frac{\partial_x^2 F}{ 8 F} \right)^2 \frac{2\hbar}{m} . \label{raw2}
\end{eqnarray}
We now note that the quantity $\bar{\bar{s}}$, defined as 
\begin{equation}
   \bar{\bar{s}}  \equiv \frac{m\bar{\lambda} |F|}{\hbar |\partial_x^2 F|}
\end{equation} 
is also a dimensionless action for the system in units of $\hbar$. As with Eq.(\ref{lnineq}), we have substituted the average Lyapunov exponent, $\bar{\lambda}$ into the strong inequalities.  Assuming that $\bar{\bar{s}}$ is of the same order as the dimensionless action of the system, $s$, Eqs. (\ref{raw1}) and (\ref{raw2}) become 
\begin{eqnarray}
   s  & \gg & \frac{\ln(s)}{16\sqrt{2}},  \label{paw1} 
\end{eqnarray}
and 
\begin{eqnarray}
    s  & \gg & \frac{\sqrt{\ln(s)}}{8} . \label{paw2}
\end{eqnarray}
These inequalities are automatically satisfied in the semiclassical regime, where $s \gg 1$.  Significantly, they will be satisfied when the semiclassical criteria given by Eq.(\ref{eq::wtos0}) is met.  The conclusion is that Eq.(\ref{eq::wtos0}) defines a semiclassical regime where the strong QCT will be satisfied when the weak QCT occurs in the Eq.(\ref{weak6}) regime.

This also constrains the time, $t_{qc}$, at which the weak QCT occurs.  Since $D=\hbar^2 k$, we can write $t_{qc}\approx m\bar{\lambda}/\hbar k$. The strong QCT will occur within the large $k$ region.  Since $k$ will be large the weak QCT will also occur quickly.  This is not surprising, as localization at a level which allows a trajectory picture should imply that interference is rapidly eliminated and a local classical picture should emerge regardless of whether the system achieves a global steady-state. 

We now turn to the question of what it means for the quantum-to-classical transition to occur in the weak sense without having occurred in the strong sense, particularly in the $k$ range we have been discussing.  It is clear that this should not happen in the low noise regime. In this regime the wave-function of an observed trajectory is small compared to the available phase space, and thus fine details in the structure of the phase-space densities are visible. The trajectories are smooth, since the noise is small in comparison to the deterministic classical dynamics. It is also clear, as mentioned above, that the trajectories must obey the classical equations of motion.  If this were false, they would not give the same fine structure as the classical density when their (well-localized) Wigner functions are averaged over all noisy realizations. It is similarly clear that when the low-noise inequalities are violated the
weak transition should be able to occur without the strong transition, as the lack of a weak noise threshold implies.  This is
because the dynamics due to noise alone is the same in both
quantum and classical systems.  Thus if noise dominates the
dynamics, then the quantum and classical densities will agree closely,
even though the observed trajectories will also be noise dominated and
will therefore not follow smooth motion of the classical
Hamiltonian. 

What is not so clear is how the weak transition 
occurs when noise does not swamp the deterministic dynamics, but
the wave-function of an observed trajectory is sufficiently
delocalized that the dynamics of its centroid remain noisy. Note that
in this case the noise on the centroid is not purely a result of the
noise introduced by the measurement/environment, but is due in large
part to the fact that the wave-function is broad. The implication is that the observer does not know well the location of the system in phase space, and thus the centroid of the wave-function changes significantly as the observer obtains the random stream of measurement results. This is what one would expect in a weaker noise domain.

The question of how the weak QCT is satisfied while violating the strong was
discussed briefly in~\cite{Greenbaum05}. We now provide more detailed results on this question, by simulating the Duffing oscillator with the same parameters as considered in~\cite{Greenbaum05}.  The Hamiltonian of the Duffing oscillator is~\cite{Lin90}
\begin{equation}
  H =  p^2/(2m) - \alpha x^2 + \beta x^4 + \Lambda \cos(\omega t)
\end{equation}
where the parameters are chosen to be $(m,\alpha,\beta,\Lambda,\omega) = (1,10,0.5,10,6.07)$, and for the quantum simulation we choose $\hbar = 0.1$. Choosing the value of $\hbar$ is merely a convenient means of setting the action of the system relative to $\hbar$. Here we fix the action (equivalently the available phase-space area $A$), and choose the area that a minimum uncertainty wave-packet occupies by setting the value of $\hbar$.  

In~\cite{Greenbaum05} the weak QCT is demonstrated for the momentum diffusion rate $D=0.01$. We now examine the behavior of the observed trajectories in this regime. The environment considered in~\cite{GreenbaumNew,Greenbaum05} is equivalent to a continuous measurement of the oscillator position, $x$, and the measurement strength $k = D/\hbar^2 = 1$. The equation of motion for the system density matrix under this continuous measurement is given by the stochastic master equation~\cite{JacobsSteck06}
\begin{eqnarray}
  d \rho & = & -(i/\hbar) [H ,\rho] dt - k [x,[x,\rho]] dt \nonumber \\ 
  & & + \sqrt{2k}(x\rho + \rho x - 2 \langle x \rangle \rho) dW 
\end{eqnarray}
where $dW$ is the increment of Wiener noise satisfying $(dW)^2 =
dt$. We choose the initial state to be a minimum uncertainty
(coherent) state with centroid $(\langle x \rangle,\langle p \rangle)
= (-3,8)$, and position and momentum variances equal to $\hbar/2 =
0.05$. The accessible phase space for the classical system has
position boundaries at approximately $\pm 5$, and momentum boundaries
at $\pm 20$. 

\begin{figure}[t] 
\leavevmode\includegraphics[width=0.98\hsize]{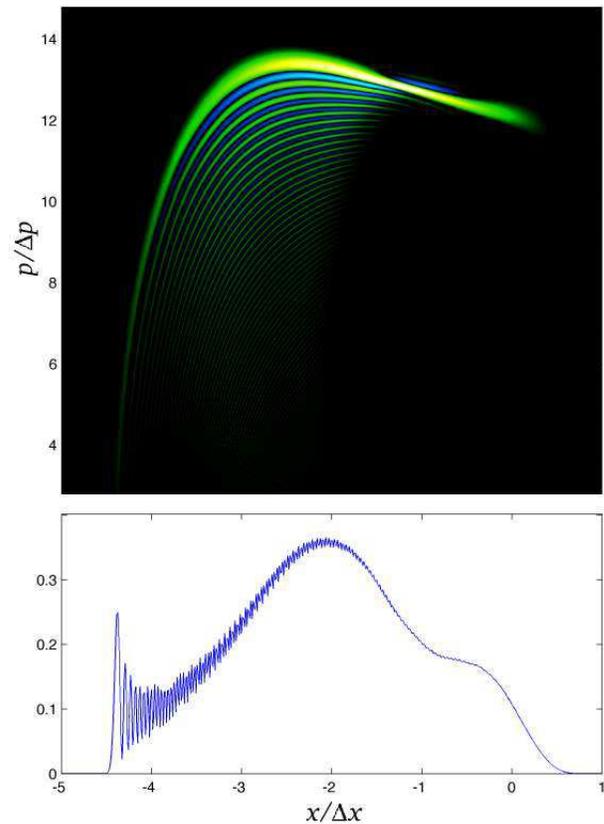}
\caption{(Color online) (a) A typical Wigner function for the Duffing oscillator when $\hbar=0.1$ and the measurement strength $k=1$. In this plot luminosity denotes the absolute value of the real part of the Wigner function (thus black corresponds to zero). (b) The associated probability density for the position of the oscillator. The wave-function is spread over a significant region of the phase-space.} 
\label{fig1}
\end{figure}
\begin{figure}[h] 
\leavevmode\includegraphics[width=0.98\hsize]{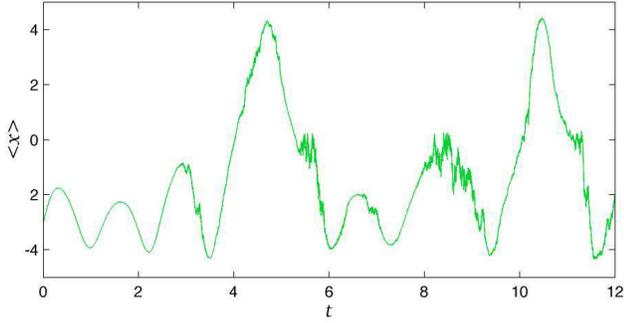}
\caption{(Color online) The mean position of the observed Duffing oscillator when $\hbar=0.1$ and $k=1$. The position uncertainty of the quantum state is manifest in the noise that is visible on this trajectory.} 
\label{fig2}
\end{figure}

 
In Fig.~\ref{fig1} we show the Wigner function for the oscillator after a time of $t=12$, (approximately $12$ periods of the drive), along with the corresponding probability density for the position of the oscillator. The position wave-function is spread over a significant region of the available phase space, and one therefore expects the trajectory for the mean position to experience significant noise. In Fig~\ref{fig2} we plot the mean position up to $t=12$, and indeed the effect of the noise is clearly visible. The quantum and classical phase-space densities can thus agree on intermediate scales and achieve a weak QCT, even if the observed trajectories do not follow the smooth classical dynamics. 

\section{Deriving the Strong Transition from the Weak}
\label{sec::IV}

In the previous section we derived the regime of the weak QCT where the strong QCT is also satisfied (Eq.(\ref{eq::wtos0})). Here we use an alternative approach to derive a set of conditions under which the weak transition will imply the strong.  To begin we note that the existence of fine structure at the scale of the phase-space area $l^2$ bounds the width of the wave-functions of the trajectories. This is because the phase space density is an average over the wave-functions of all trajectories, and this automatically precludes the phase-space density from having oscillations smaller than the width of the wave-function.  Using $ (m |\partial_x F|)^{1/4} = \sqrt{m \lambda}$ as the scaling factor between position and the phase-space length $l$, this bound is  
\begin{equation}
  V_x \leq m \lambda l^2 . 
  \label{Vxl}
\end{equation} 
Thus if $l$ is small enough, then it will force the wave-function for
the corresponding trajectory to be localized. This, in turn, will force it to satisfy the conditions of the strong QCT. In this case the weak transition will imply the strong. This is because all three strong inequalities, Eq.(\ref{locineq2}) (or Eq.(\ref{locineq3})), and Eq.(\ref{lnineq}) are in fact a result of conditions limiting the position variance, as shown in Ref.(~\cite{Bhattacharya03}). We can therefore derive quantitative inequalities determining when the weak QCT will imply the strong, by using the strong bounds on $V_x$, then Eq.(\ref{Vxl}) to bound $l$, and finally Eq.(\ref{weak1}) to derive bounds on $k$. 

There are three strong bounds on $V_x$. The bound that leads to the localization condition (Eq.(\ref{locineq2}) or (\ref{locineq3})) and the two bounds that lead respectively to the two low noise inequalities given in Eq.(\ref{lnineq}). The bound that leads to the localization inequality is~\cite{Bhattacharya03} 
\begin{equation}
      V_x \ll \frac{2F}{\partial_x^2 F}
\end{equation}
Using the procedure just described, this gives the following condition on $k$: 
\begin{eqnarray}
  k & \ll &  \left( \frac{4 m\lambda^2}{\hbar} \right) \left[ \frac{\bar{\bar{s}}}{\ln[\bar{s}\tilde{s} / (2 \bar{\bar{s}})]}  \right] \nonumber \\ 
     & \approx & \left( \frac{ 4m\lambda^2}{\hbar} \right) \left[ \frac{s}{\ln(s/2)} \right] \label{eq::wtos1} . 
\end{eqnarray} 
The bound on $V_x$ that leads to the left hand side of Eq.(\ref{lnineq}) is $V_x \ll  \sqrt{S/(k m)}$~\cite{Bhattacharya03} where $S$ has units of action and is given by Eq.(\ref{eq:s1}). This leads initially to the inequality 
\begin{equation}
    k  \ll  \left( \frac{2 m \bar{\lambda}}{\hbar} \right) \left( \frac{S}{2 \hbar} \right)^{1/3} \ln\left( \xi A \sqrt{\frac{k}{S m \bar{\lambda}^2}}\right)^{-1} 
\end{equation}
To complete the derivation we need to eliminate $k$ from the right hand side. Since we are deriving a condition for when the weak transition implies the strong, we can assume that the weak QCT takes place in the regime given by Eq.(\ref{weak6}).  So as to be conservative (that is, to derive the weakest condition) we should choose the value of $k$ on the right hand side to be as large as possible. A very conservative value for $k$ is to saturate the upper bound in Eq.(\ref{weak6}), and this gives
\begin{eqnarray}
    k & \ll & \left( \frac{2 m \bar{\lambda}}{\hbar} \right) \left( \frac{S}{2 \hbar} \right)^{1/3} \ln\left( \bar{s} \sqrt{\frac{ 2\bar{s} \tilde{s}}{\ln\tilde{s}}}\right)^{-1} \nonumber \\ 
      & \approx &  \left( \frac{2 m \bar{\lambda}}{\hbar} \right) 2^{2/3}  \left[ \frac{s^{1/3}}{ \ln( 2 s^4/ \ln s )  } \right] \label{eq::wtos2}
\end{eqnarray}
The third and final bound on $V_x$ is~\cite{Bhattacharya03}
\begin{equation}
  V_x^2 \ll \frac{1}{4 m k} \sqrt{\frac{m |F|^3}{8k |p \partial_x F|}} 
\end{equation}
This results in the condition 
\begin{equation}
  k \ll   \left( \frac{2 m \bar{\lambda}}{\hbar} \right) 8^{-1/7}  \left[ \frac{s^{1/7}}{ \ln( 32 s^5 [\ln s]^{-3/2} )  } \right] , \label{eq::wtos3}
\end{equation}
where we have assumed that $S'/\hbar \approx s$. 

If we satisfy each of the three inequalities given by Eqs.(\ref{eq::wtos1}), (\ref{eq::wtos2}) and (\ref{eq::wtos3}), then the weak transition will imply the strong transition.  The second and third conditions, Eqs.(\ref{eq::wtos2}) and (\ref{eq::wtos3}) are, however, more stringent than the weak regime we have invoked. In order be within the localized regime and satisfy these conditions, we must at least have 
\begin{equation}
  \frac{1}{\ln s} \ll  \left[ \frac{s^{1/7}}{ \ln( 32 s^5 [\ln s]^{-3/2} )  } \right]  . \label{eq::wtos4}
\end{equation}
When $s\geq 10$, the denominator on the right hand side is well approximated by $1/(5\ln s)$, and so we have  
\begin{equation}
 s \gg  5^7 \approx 10^5 
\end{equation}
Comparing this with the equivalent condition in
Section~\ref{sec::III}, Eq.(\ref{eq::wtos0}), we see that while the two results are similar, the new result is well above the threshold set in Section III. Thus the above analysis, while providing an alternative approach, reinforces the interpretation of that section.  We may therefore conclude that the criteria given by Eq.(\ref{eq::wtos0}), derived indirectly by comparing the weak and strong QCT, will also be met by the more intuitive criterion derived in this section.
\vspace{6mm}

\section{Conclusion} 
\label{sec::V}

There are two ways to ask if (nonlinear) classical dynamics has emerged from the evolution of a quantum system. One is to observe the system and to ask when the motion of a localized centroid is indistinguishable from the classical trajectories. When this is true we refer to the system as having made the transition in the strong sense. The other method is to obtain only the phase-space probability densities for the classical and quantum motion, and to ask when the these densities become indistinguishable. When this is true we say that the system has made the transition in the weak sense. Two distinct methods have been used to determine how an open quantum system will make the transition to classical dynamics. 

Here we have shown that in the semiclassical regime (the regime in
which the weak inequalities are valid), these two levels of description may be compared.  Specifically, when the action of the system is much larger
than $\hbar$, the inequalities implying a rapid weak QCT, which takes place before a classical steady-state, are
stronger than those implying the strong QCT.  We have also pointed out that when the action is much larger than $\hbar$, and the environmental noise is very small, both this weak regime and the strong transition are essentially equivalent, regardless of the exact behavior of the respective inequalities.  

From the above analysis we have also shown that in the semiclassical
regime the strong inequalities may be simplified, so that in both the
weak and strong cases, the conditions for the emergence of classical
motion involve simple inequalities. The inequalities accompanying the strong QCT being 
\begin{equation}
\frac{1}{s} \; \ll \;  \frac{D}{2 \hbar m \bar\lambda^2}  \; \ll \;   \frac{s}{8}
\end{equation}
while, in defining this weak region where the QCT precedes the termination of classical structure, we get 
\begin{equation}
  \frac{1}{\ln s} \; \lesssim \;  \frac{D}{2 \hbar m \bar{\lambda}^2 } \; \ll \;  \frac{s}{\ln s}.    \label{weak6new}
\end{equation}
Here $D$ is the momentum diffusion coefficient due to the measurement or environment, $\lambda$ is the Lyapunov exponent for the system, $s$ is the action of the system in units of $\hbar$, and $m$ is the mass. 

We have also derived a very simple sufficient condition for when this weak regime implies the strong transition, and this is $S/\hbar \gg 10^3$, where $S$ is the action of the system.  When this condition is not met, the weak transition occurs without the smooth trajectories of classical mechanics. However, in the semiclassical limit this weak regime is entirely sufficient to determine the emergence of classical dynamics in a quantum system. 


\begin{thebibliography}{27}
\bibitem{foot1}Specifically, this averaging process involves calculating the Wigner function for each trajectory, and then summing these Wigner functions with each weighted by the probability that the corresponding trajectory occurs.
\end{thebibliography}

\end{document}